\documentclass[superscriptaddress,showpacs,twocolumn,aps,pre,10pt]{revtex4-1}
\usepackage{natbib}
\usepackage{graphicx,dcolumn,bm,bbm,amsmath,amssymb,color}

\newcommand{\bhu}{ \hat{\bf u} }
\newcommand{\br}{ {\bf r} }
\newcommand{\bv}{ {\bf v} }

\begin{document}
\title{Active dipole clusters: from helical motion to fission}

\affiliation{Institut f\"ur Theoretische Physik II: Weiche Materie,
Heinrich-Heine-Universit\"at D\"{u}sseldorf,
Universit{\"a}tsstra{\ss}e 1, 40225 D\"{u}sseldorf, Germany}

\author{Andreas Kaiser}
\email{kaiser@thphy.uni-duesseldorf.de}

\author{Katarina Popowa}

\author{Hartmut L\"owen}

\date{\today}


\begin{abstract}
The structure of a finite particle cluster
is typically determined by total energy minimization. Here
we consider the case where a 
cluster of soft sphere dipoles becomes active, i.e. when
the individual particles exhibit an additional self-propulsion along their dipole moments.
We numerically solve the overdamped equations of motion  for
soft-sphere dipoles in a solvent.
Starting from an initial metastable dipolar cluster, the self-propulsion generates
a complex cluster dynamics. The  final cluster state has in general a structure widely different
to the initial one, the details depend on the model parameters and on the protocol of how the self-propulsion is turned on.
The center-of-mass of the cluster moves on a  helical path, the details of which 
are governed by the initial cluster magnetization.
An  instantaneous switch to a high self-propulsion leads to fission of the cluster.
However, fission does not occur if the self-propulsion is increased slowly  to high strengths.
Our predictions can be verified through
experiments with self-phoretic colloidal Janus-particles and for macroscopic self-propelled dipoles in a highly viscous solvent.
\end{abstract}

\maketitle

\section{Introduction}
\label{sec:intro}

Clusters comprising $N$ individual particles occur in widely different areas of physics ranging from the atomic world
~\cite{Wales_review} to nanoparticles~\cite{review_nanoclusters},  colloids~\cite{review_colloids,Pine_Science_2003,Glotzer,Kraft_Wittkowski_PRE_2013,Nemeth_Lowen_JPCM_1998,RoyallJPCM2009,Wales_Nano12,BrennerPNAS2011,Grzybowski}
and to macroscopic granulates~\cite{granular_clusters}. In the simplest case, the particles interact via a
pairwise potential, such as a Lennard-Jones potential~\cite{References_pure_LJ} or a hard-sphere-dipole 
interaction~\cite{DietrichPRE94,Miller,Farrell,Patey,Holm,Kantorovic,Klapp,Messina_DipoleCluster,RehbergArxiv}, and the equilibrium
groundstate structure of the cluster is obtained by minimization of the total potential energy.
Even for small $N$ the structure can be nontrivial and differs substantially from a cutout of a simple crystal.
Here, we consider clusters where the constituents are active or self-propelled particles
which are swimming in a viscous solvent at low Reynolds number.
Such active particles or microswimmers can be artificially realized as colloidal Janus particles
exposed to a thermal gradient~\cite{Sano_PRL2010} close to a solvent phase transition~\cite{Bechinger} or a
chemical reactant catalyzed at one part of the Janus 
particle~\cite{Paxton,Baraban_ACSnano,Kaiser_PRE2013Janus,SenPNAS13,BocquetPRL2010,Takagi,Poon_SM14,Ebbens_EPL14,DietrichSwimmer}.
This brings the particle into motion along its symmetry axis, thereby creating a nonequilibrium situation 
where the minimization principle of the potential energy breaks down.
An ``active cluster'' composed of aggregated self-propelled
particles will therefore exhibit a characteristic structure, and a characteristic motion. 
In particular, as the self-propulsion force grows stronger, the cluster can break revealing an
 activity-induced fission.
The occurrence of clusters within active suspensions has been studied frequently in the last years through experiments~\cite{Poon,Bialke_PRL2013,Palacci_science,BocquetPRL12},
theory~\cite{CatesClusterTheory,BialkeReview}, and simulations~\cite{MarchettiPRL12,ZoettelStarkPRL14,CR_SM14,PohlStarkPRL14} 
considering purely repulsive as well as slightly attractive particle interactions~\cite{BaskaranPRE13}.
However, controlled fission of such clusters has not been a focus of research, with first forrays in this field made 
only recently by Soto and Golestanian~\cite{Soto_Golestanian_PRL_2014,Soto_Golestanian_2015}.

We describe these phenomena by a simple model combining 
Weeks-Chandler-Andersen and dipole pair potentials with an intrinsic effective self-propulsion 
force~\cite{BtHNatComm,BtHJPCM2014} along the dipole moment~\cite{KlappSwimmer}. 
Our model differs from the one employed by Soto and Golestanian~\cite{Soto_Golestanian_PRL_2014,Soto_Golestanian_2015}, 
which focuses on clusters of self-propelled particles induced by chemotaxis and characterized by their violation of the actio-reactio principle.

In order to keep the model as simple as possible, we consider in this paper only a small number $N$ of dipoles
for the trivial cases $N=1,2$ up to $N=5$. The particles are soft spheres with an embedded dipole 
moment that is oriented along the direction of self-propulsion. The situation with $N=4$ and $N=5$ dipoles already reveals 
a quite complex dynamical behaviour.
Our initial configuration is either a linear or ring-like structure which represents the groundstate of the 
system~\cite{Miller,Messina_DipoleCluster,Kantorovic} or a metastable structure such as various compact and branched 
clusters, in particular a Y-junction of $N=4$ dipoles~\cite{SafranNatMat,Rovigatti}.
We then introduce the self-propulsion and follow the dynamics of the particles. When doing so,
 we distinguish between two different protocols: an instantaneous switching to a given velocity and an ``adiabatic'' 
slow increase of the self-propulsion. 
As a result, we find that the cluster's center-of-mass generally moves along a helical trajectory.	 
The details of the helix are governed by the magnetization of the initial cluster.
Moreover, we find a plethora of final cluster states, which depend on the model parameters and the enacted protocol.
For strong self-propulsion and a nonlinear initial state
with nonvanishing dipole moment, fission of the cluster occurs for instantaneous switching. However, interestingly,
for the same parameters, there is no fission in the adiabatic switching case.

Our results can be verified for colloidal Janus particles which a strong dipole moment
along their symmetry axis. At vanishing self-propulsion
these particles have been considered quite a lot in the context of ferrogels and 
ferrofluids~\cite{Patey,Holm,Holm2,Kantorovic,Klapp,Messina_DipoleCluster}. Dipolar self-propelled particles have been
prepared recently by Baraban and coworkers~\cite{Baraban_Nano2013,Baraban_ACSNano2013} and aggregation into clusters has indeed
been observed~\cite{LB_pc}.	

Our paper is organized as follows: In Sec.~\ref{Sec:model} we specify our model for active dipolar particles.
The different types of motion of clusters with and without vanishing total dipole moment are presented in Sec.~\ref{Sec:Trajectories}. In Sec.~\ref{Sec:Clusters} we study the fission induced by activity in detail and we conclude in Sec.~\ref{Sec:conc}.

\section{Model}
\label{Sec:model}

\begin{figure}[tbhp]
\begin{center}
\includegraphics[clip=,width= \columnwidth]{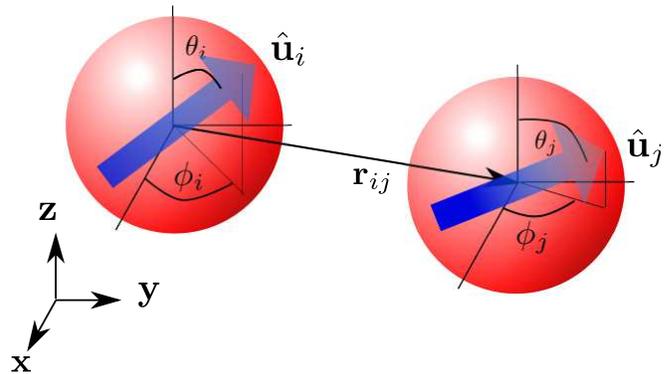}
\caption{ \label{f1} Sketch of a pair of dipoles in three spatial dimensions with center-of-mass distance $\br_{ij}$. 
The self-propulsion and the dipole moment are directed along the unit vector 
$\bhu_i= (\sin \theta_i \cos \phi_i, \sin\theta_i \sin \phi_i, \cos \theta_i)$.}
\end{center}
\end{figure}

We consider $N$ spherical dipoles in three spatial dimensions. The position of the $i$th dipole will be denoted 
as $\br_i=[x_i,y_i,z_i]$ and 
the dipole moment ${\bf m}_i=m\bhu_i$ is directed along the unit vector 
$\bhu_i  = (\sin \theta_i \cos \phi_i, \sin\theta_i \sin \phi_i, \cos \theta_i)$.
The total pairwise interaction potential $U_{ij}$ is the sum of a
Weeks-Chandler-Andersen $U^{\text{WCA}}$, which %is a truncated and shifted Lennard-Jones potential and
describes a repulsive soft core~\cite{WCA} and a point dipole potential $U^{\text{D}}$ 
\begin{eqnarray}
U^{\text{WCA}}_{ij} &=& \begin{cases} 4\epsilon \left[  \left( \frac{\sigma}{r_{ij}} \right)^{12} - \left( \frac{\sigma}{r_{ij}} \right)^6 \right] + \epsilon, &r \leq r_c,\\0,  &r > r_c, \end{cases}\\
U^{D}_{ij} &=& \frac{m^2}{r_{ij}^3} \left[ \bhu_i \cdot \bhu_j - \frac{3 (\bhu_i \cdot \br_{ij}) (\bhu_j \cdot \br_{ij})}{r_{ij}^2}  \right],
\label{eq:Potential}
\end{eqnarray}
where $\br_{ij}=\br_i-\br_j$ is the position of particle $j$ relative to particle $i$ and $r_{ij}$ the respective 
distance, see Fig.~\ref{f1}. 
Here, the cut-off length is $r_{c} = 2 ^{1/6} \sigma$.
We introduce the self-propulsion by means of an effective internal force ${\bf F}_i=F\bhu_i$~\cite{BtHComment,BtHJPCM2014}
which is directed along all dipole moments ${\bf m}_i$, leading to a constant self-propulsion velocity 
$\bv_i = v \bhu_i$ for an individual single particle. The velocity is given by $v=F/f_t$, and $f_t$ denotes the 
translational Stokes friction coefficient.
As units of energy and length we choose the parameters  $\epsilon$ and  $\sigma$ from the WCA-potential. 
Time is measured in units of $\tau = f_t \sigma^2 / \epsilon$.

The motion of microswimmers is restricted to the low Reynolds number regime and the corresponding overdamped
equations of motion for the positions $\br_i$ and orientations $\bhu_i$ are 

\begin{eqnarray}
f_t \cdot \partial_t \br_i (t) &=& -\nabla_{{\br}_i} U+ f_t v \bhu_i (t),\\
f_r \cdot \partial_t \bhu_i (t) &=& - {\bf T}_i \times \bhu_i (t),  
\label{eq:EOMs}
\end{eqnarray}
where ${\bf T}_i = \bhu_i (t) \times \nabla_{\bhu_i} U$ is the torque on particle $i$ 
and $U = 1/2 \sum_{i,j\neq i} {U_{ij}}$ the total interaction potential. The 
rotational friction coefficient $f_r$ of the spherical particles 
is assumed to be linked to $f_t$ via the equilibrium relation $f_r = f_t \sigma^2 / 3$. 
We neglect any thermal fluctuations and any solvent-induced hydrodynamic interactions between the particles in our model.

Simulations are performed in three spatial dimensions using a simple Euler integration
algorithm with time steps $\Delta t = 10^{-4} \tau$ for simulation times $t=400\tau$, 
which are sufficiently large enough to allow the active dipole clusters to achieve steady state structures.
The dipole strength is varied in the range $0 \leq m^2 / (\epsilon \sigma^3) < 6$.

Starting configurations are gathered by an energy minimization for the respective parameter sets for passive dipoles, $v=0$.
We consider two protocols of how the activity is applied to the dipoles. 
The self-propulsion is either instantaneously increased 
for the starting configurations, or we slowly increase the activity 
stepwise by a velocity increase of $\Delta v =0.1 \sigma / \tau$
subsequent after a long waiting time of $t=400\tau$.
This corresponds to slow or 'adiabatic' switching.

\section{Active dipole clusters: Simple cases and center-of-mass trajectories}
\label{Sec:Trajectories}

\subsection{The special cases $N=1,2,3$}

Let us now discuss the simple cases of very few particles $N=1,2,3$. First of all, a single particle $N=1$ will trivially
move with velocity $v$ on a straight line along its orientation $\hat {\bf u}$.
Next, the ground state of two dipoles ($N=2$) is a head-tail-configuration where the two magnetic moments possess the same 
orientational direction and are aligned along the separation vector of the two spheres. This is shown in 
Fig.~\ref{Trajectories}(c). Putting a drive along the magnetic moments will lead to joint motion of this linear cluster
with the same velocity $v$ as that for the individual spheres. This behaviour does not depend on the protocol of the drive.

Third, a magnetic triplet ($N=3$) will form a metastable ring-like structure where the sphere centers are on an equilateral triangle
and the orientations have relative angle differences of $2\pi/3$, see Fig.~\ref{Trajectories}(b). However, the
groundstate is a linear chain~\cite{Messina_DipoleCluster}. This ring-like cluster will rotate
around its non-moving center-of-mass when a self-propulsion is turned on. The distance between the spheres and the angular 
velocity increase with increasing drive.

\subsection{Center-of-mass trajectories}

\begin{figure}[thbp]
\begin{center}
\includegraphics[clip=,width=1\columnwidth]{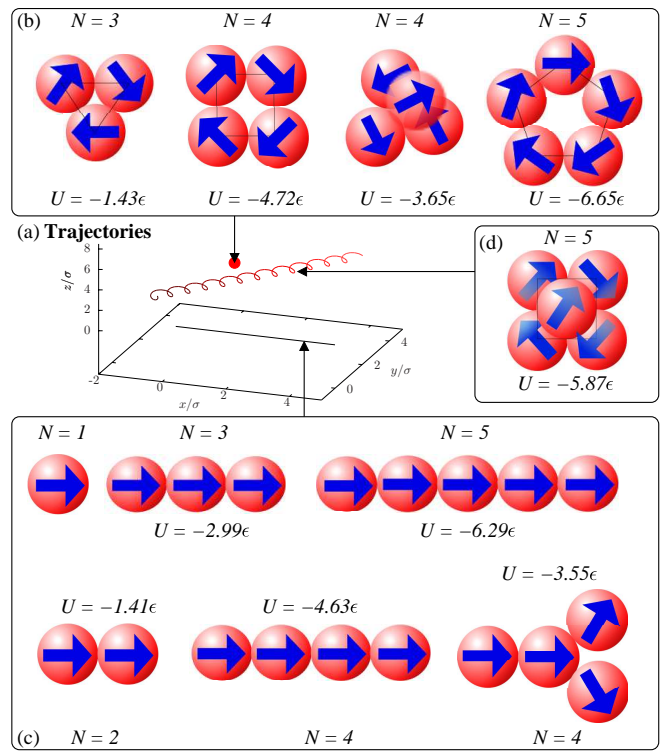}%{TrajAD}
\caption{ \label{Trajectories} (a) Trajectories of the center-of-mass for different initial 
active dipole clusters composed of $N=1,\ldots,5$ dipoles. 
(b) Ring-like clusters perform a rotation around the center-of-mass such that the center-of-mass is non-moving.
(c) Chains, as well as a Y-junction, move along a straight line and (d) a compact cluster of $N=5$ dipoles propagates 
on a helix. Please note that the relative positions of the cluster particles change during the cluster propagation 
for an initial Y-junction and the compact cluster shown in~(d). For all cluster conformations the respective total potential energy $U/\epsilon$ is given for $m^2/(\epsilon\sigma^3) = 1$.}
\end{center}
\end{figure}

Next, we consider the center-of-mass coordinate of dipole clusters 
as defined by
\begin{eqnarray}
{\bf R}_c (t)= \frac{1}{N}\sum_{i=1}^N \br_i (t)\,.
\label{eq:Rcom}
\end{eqnarray}
Due to the reciprocal interactions between the individual dipoles, the motion of the center-of-mass ${\bf R}_c$ 
can be determined by the orientations $\bhu_i$ of the active particles,
\begin{eqnarray}
\partial_t{\bf R}_c (t)= \frac{1}{N}\sum_{i=1}^N \partial_t\br_i (t) =\frac{f_tv}{N}\sum_{i=1}^N \bhu_i(t) \,,
\label{eq:eomRc}
\end{eqnarray} 
such that the center-of-mass velocity $v_c$ is proportional to the total
magnetic moment
\begin{eqnarray}
M=  \left| \sum_{i=1}^{N} {\bf m}_i \right| = m \left| \sum_{i=1}^{N} \bhu_i \right| \,.
\label{eq:dipmom}
\end{eqnarray}
The motion of the center-of-mass can be easily classified as summarized in Fig.~\ref{Trajectories}.
The simplest situation is an initial {\it ring cluster\/} as shown for $N=3,4,5$ in Fig.~\ref{Trajectories}(b)
whose center-of-mass is in the ring center. This ring has a vanishing total dipole moment, $M=0$.
Based on a simple analysis of the equation of motions and on symmetry arguments, 
an initial ring cluster will just rotate around its center-of-mass such that the center-of-mass is non-moving.
This is valid for any $N$, for any self-propulsion strength $v$ and for any protocol (similar to the 
special case $N=3$ discussed above)~\cite{SuppMat}. Note, however, 
that the radius of the rotating ring and the corresponding angular velocity do increase with increasing drive $v$.
The result of a non-moving center-of-mass can also be obtained for other initial clusters with vanishing initial $M$.
An example for a compact three-dimensional cluster different from a ring is shown in Fig.~\ref{Trajectories}(b) for $N=4$.
This tetrahedral cluster is mechanically stable (but energetically metastable) 
in equilibrium $(v=0)$ and has no spontaneous magnetization $(M=0)$.

Obviously, linear chains of $N \geq 1$ dipoles move on a straight line and do not change their relative shape, 
see Fig.~\ref{Trajectories}(c), 
since all dipole moments and consequently the self-propulsion
velocities are directed along the main symmetry axis.
There are more complex structures
like a Y-junction with $N=4$ dipoles, see again Fig.~\ref{Trajectories}(c), whose center-of-mass moves on a  straight line.
The relative positions of these cluster particles, however, can change. The details of the relative motion 
will be considered in the next section.
Finally, a moving cluster of constant shape generally performs a helical motion~\cite{WittkowskiPRE12}.
The helical motion is generated by a non-vanishing magnetization $M\not= 0$ which provides 
the translational force and a non-vanishing torque on the cluster. A typical example 
is an initially compact cluster with non-vanishing magnetization as shown 
for $N=5$ particles in Fig.~\ref{Trajectories}(d). This compact cluster has the dipoles arranged at the corners of a pyramid.

\section{Reorganization and fission of active dipole clusters}
\label{Sec:Clusters}

We now turn to clusters composed of four ($N=4$) and five ($N=5$) particles. These cases reveal nontrivial and interesting 
dynamical behaviour caused by self-propulsion. In particular, some structures will reorganize, reassemble and split due to the
self-propulsion and this depends explicitly on the protocol applied to turn on the self-propulsion. To be specific we first consider 
an initial metastable Y-junction for $N=4$ particles and then study a compact metastable initial cluster for $N=5$ particles.

\subsection{Y-junction $(N=4)$}

\begin{figure*}[tbh]
\begin{center}
\includegraphics[clip=,width=2\columnwidth]{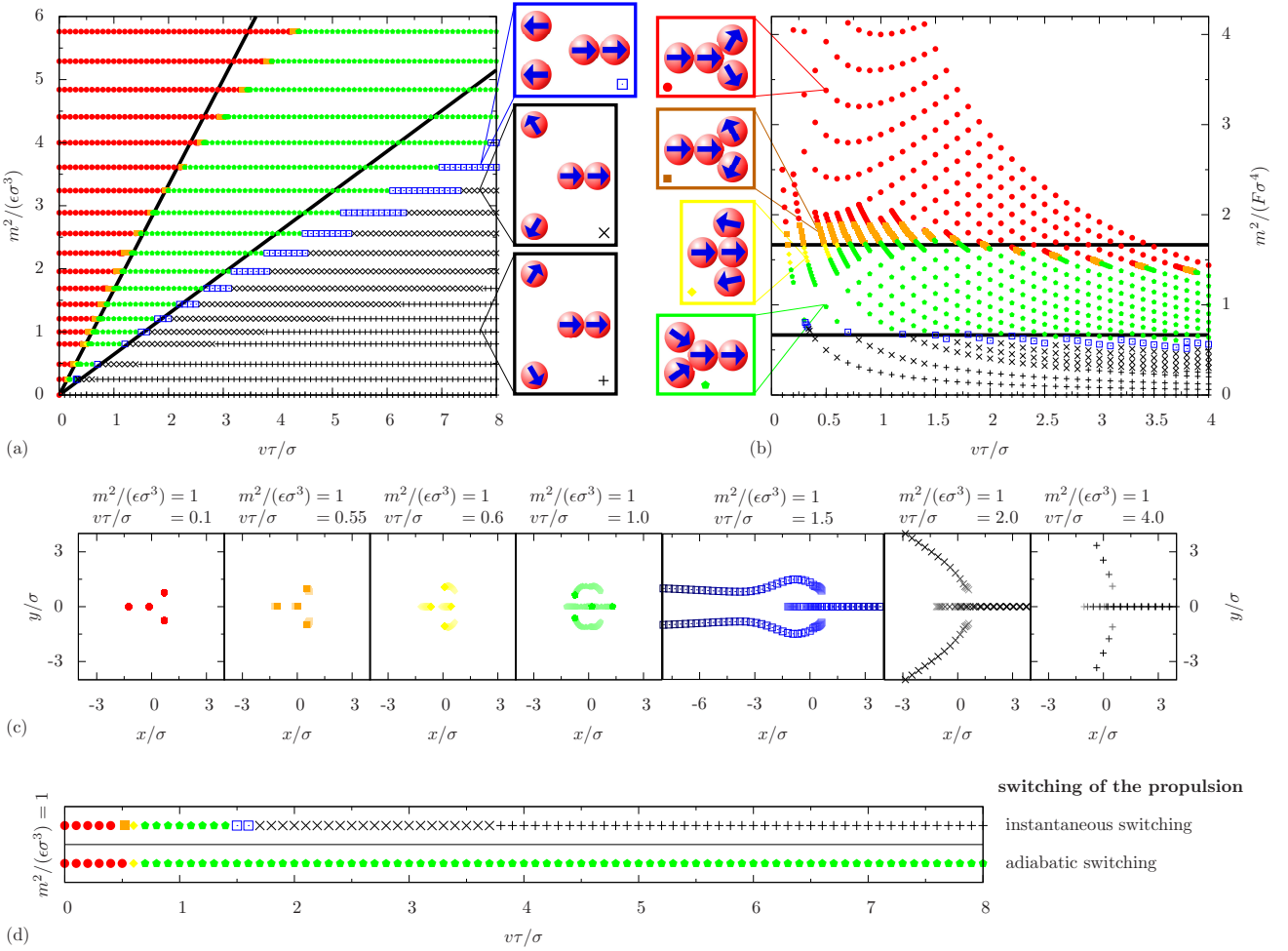}
\caption{ \label{PhaseDiag} (a) Emerging state diagram for an initial Y-junction for varied reduced dipole  
strength $m^2/(\epsilon \sigma^3)$ and  self-propulsion velocity $v$. Each symbol represents 
a different type of the final cluster configuration, as shown by snapshots (middle). 
The black lines indicate an ideal hard sphere scaling. (b) Rescaled state diagram, showing that state transitions are 
proportional to $m^2/(F\sigma^4)$ for small self-propulsion and dipole strengths. (c) Trajectories of the individual dipoles 
 observed in the moving center-of-mass frame~\cite{SuppMat}.
(d) Comparison of the emerging clusters for fixed reduced dipole strength $m^2/(\epsilon \sigma^3)$ and varied reduced 
self-propulsion $v\tau/\sigma$ for an instantaneous and an adiabatic switching protocol.}
\end{center}
\end{figure*}

\begin{figure}[thbp]
\begin{center}
\includegraphics[clip=,width=\columnwidth]{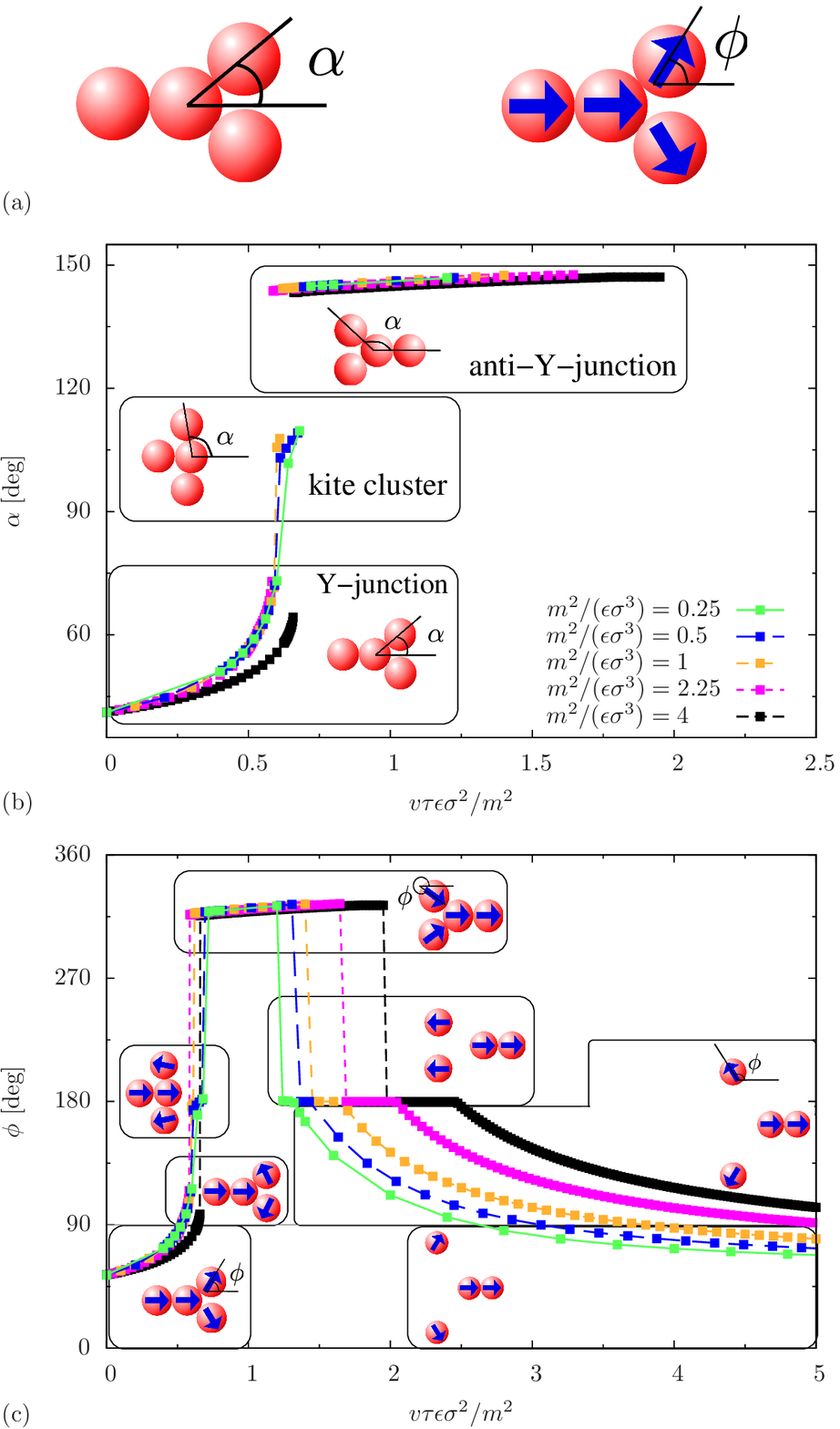}
\caption{ \label{angles} (a) Schematic representation of the final angles $\alpha$ and $\phi$
and (b),(c) their dependence of the self-propulsion velocity $v$, here rescaled with the fixed dipole strength. 
Respective conformations characterized by the order parameters are indicated by sketches.}
\end{center}
\end{figure}

An initial metastable Y-junction built up by $N=4$ dipoles is shown in Fig.~\ref{Trajectories}(c). 
We study its dynamical behaviour for varying dipole strength, self-propulsion velocity $v$ and switching protocol. In this case
the motion is two-dimensional in the plane set by the initial configuration.

Figure~\ref{PhaseDiag}(a) shows the cluster state diagram after a long simulation time, $t=400\tau$, in the two-dimensional 
parameter space of dipole strength $m^2$ and an instantaneously applied self-propulsion velocity $v$.
As a result, seven different final configurations can be discriminated by suitable order parameters, as explained in detail below. 
Clearly, for small self-propulsion the cluster retains its shape as it is metastable but exhibits a drift along its 
total magnetization {\bf $M$}. This is indicated by the (red) filled circle. Obviously one finds this situation 
when the self-propulsion velocity is small. For very large drives $v$, on the other hand, {\it fission\/} of the cluster shows up
(black plus symbols) where the separating dipoles fly 
away  pointing  forward. In between these two extremes, there are five other states: 
i) a Y-junction where the orientations of the 'head dipoles' now point backwards (brown filled square), 
ii) a kite cluster where these dipoles shift back towards the center of the whole cluster (yellow filled diamond), 
iii) a configuration which we denote as anti-Y-junction since 
the former 'head particles' are located in the back (green filled pentagon), 
iv) a state where two clusters occur, the backbone and two parallel 
dipoles moving in the opposite direction (blue non-filled squares) and finally another
 fission state where the separating dipoles burst away pointing backwards (black cross).
The corresponding trajectories of the dipoles during rearrangement of the clusters can be seen in Fig.~\ref{PhaseDiag}(c),
 in the center-of-mass frame of reference~\cite{SuppMat}.

It is instructive to compare the scaling behaviour of our findings with that of hard sphere dipoles. 
Since hard spheres only bear a length scale (such as the diameter $\sigma$) 
and no energy scale, the only ratio which matters, 
is that of the dipole-dipole interaction force, proportional to $m^2/\sigma^4$, relative to the internal driving force  $F$.
Therefore, in order to test this scaling, we have plotted in Fig.~\ref{PhaseDiag}(b)
the same diagram with the scaled dipolar strength $m^2/(F\sigma^4)$. If hard sphere scaling would hold, 
the transition lines should be horizontal in this plot. Deviations from this scaling then need to be attributed 
to the softness of the WCA-potential. Indeed, as is revealed in Fig.~\ref{PhaseDiag}(b), the transition 
lines between different states are almost horizontal. The largest deviation occurs for the transitions where the 
kite-like structure (yellow filled diamond) and the backward Y-junctions  (brown filled square) are involved.

Next, we change the protocol to adiabatic switching, see Fig.~\ref{PhaseDiag}(d).
By slowly increasing the self-propulsion velocity, the initial Y-junction deforms into a kite cluster. Further increase of $v$
leads to temporal detaching from the backbone and formation of an anti-Y-junction, but fission does not occur.
This can intuitively be attributed to the fact that the attraction of the dipole cluster has a larger impact on the cluster connectivity
if the particles are perturbed smoothly over time.

We now discuss the details of the different structures with the help of suitable order parameters. Let us first introduce the 
positional angle $\alpha$ of one of the 'head dipoles' position relative to the backbone of the Y-junction, see Fig.~\ref{angles}(a).
We further define an angle $\phi$ of the orientation of the 'head dipole' relative to the backbone,  see again Fig.~\ref{angles}(a).
In general, the angles $\alpha$ and $\phi$ are different.
A third order parameter is provided by the distance $d$ of the head-dipole to the next backbone particle.
We define connectivity of the cluster by the simple distance criterion: if $d<r_c$ the cluster is connected, else it is split.
Figure~\ref{angles}(b) shows the final angle $\alpha$ for fixed  dipole strength 
 and varied  self-propulsion velocity $v$ in the case of the instantaneous switching protocol.
In line with the state diagram discussed previously, for increasing $v$,
 the angle $\alpha$ broadens and the initial Y-junction $(\alpha < 90^{\circ})$ becomes a kite cluster 
($90^{\circ} \leq \alpha < 120^{\circ})$, which reorganizes into an anti-Y-junction
($120^{\circ} \leq \alpha < 180^{\circ})$. Here, the nearest neighbour is now the dipole in the rear of the backbone, 
before finally fission occurs. In the limit of small dipolar interactions, an arbitrary self-propulsion leads to fission, 
see again Fig.~\ref{PhaseDiag}(a).
The test of hard-sphere scaling, i.e.\ a scaled plot versus $m^2/(F\sigma^4)$ [see Fig.~\ref{PhaseDiag}(b)], reveals  a 
Master curve up to $m^2 / (\epsilon\sigma^3) < 3$.
This shows that the hard sphere limit $(\epsilon \rightarrow \infty)$ is reached quickly and that the
state diagram essentially depends on the parameter $F\sigma^4 / m^2 = v\tau\epsilon\sigma^2 / m^2$ for
$m^2 / (\epsilon\sigma^3) < 3$, which is consistent with Fig.~\ref{PhaseDiag}.

Likewise, in Fig.~\ref{angles}(c) the angle $\phi$ is shown, which provides a better resolution of structural details.
The initial passive Y-junction has an angle $\phi_0 = 53.4^{\circ}$. Self-propulsion first induces a broadening of $\phi$
in the Y-junction in the range $(\phi_0 \leq \phi < 90^{\circ})$ (red filled circles). Then the orientation of the
outermost particles flips to the opposite direction $(90^{\circ} \leq \phi < 180^{\circ})$  (brown filled squares)
until the kite cluster occurs, where the head dipoles turn until they are 
reversed relative to the backbone magnetization, 
$\phi \approx 180^{\circ}$. This conformation only shows up for small dipolar strengths $m^2$.
For even larger $v$,  the Y-junction temporarily splits into a remaining backbone and two 
single dipoles. These dipoles will be attracted again to the rear of the backbone,
and reattach, being orientated again in the direction of motion $(270^{\circ} < \phi < 360^{\circ})$. This reassembled structure 
is denoted as an anti-Y-junction. If $v$ is increased further, these detached dipoles may still be attracted by 
the backbone. However they may just collide in its wake and align with each other $(\phi=180^{\circ})$, leading to a 
configuration of two clusters propagating in opposing directions.
If $v$ exceeds a critical threshold,  
a fission into several units will ultimately occur.
Again, we can discriminate between a state where the detached dipoles move in the opposite 
$(90^{\circ} \leq \phi < 180^{\circ})$ or in the same direction $(\phi_0 \leq \phi < 90^{\circ})$ as the backbone of the 
initial cluster.  
As a final remark, the hard sphere scaling has a similar performance in the angle $\phi$ as shown in Fig.~\ref{angles}(c).

\begin{figure}[tbhp]
\begin{center}
\includegraphics[clip=,width= \columnwidth]{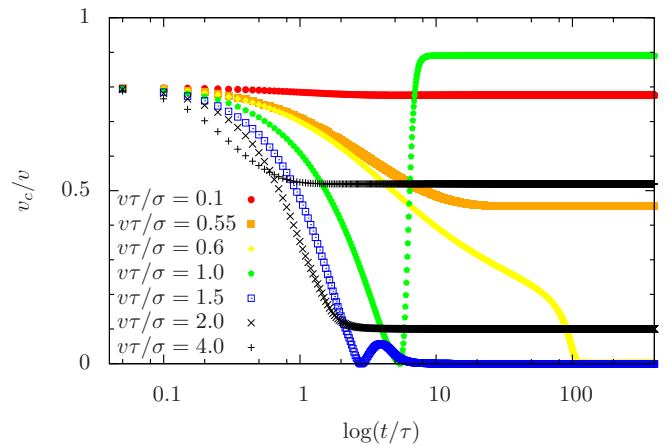}
\caption{ \label{Vcom} Reduced center-of-mass velocity $v_{c}/v$ of an initial Y-junction as a function of time
for fixed instantaneously applied self-propulsion velocity $v$ for each emerging cluster conformation using the color coding of
 Fig.~\ref{PhaseDiag} and fixed dipole strength $m^2/(\epsilon\sigma^3)=1$.}
\end{center}
\end{figure}

Finally, the center-of-mass velocity $v_c = |$d${\bf R}_c(t) / $d$t|$ (see Eq.~(\ref{eq:eomRc}))
is shown in Fig.~\ref{Vcom} for all emerging cluster conformations. 
While the kite cluster (yellow filled diamond) and the split-into-two-units state 
(blue open squares) have a vanishing center-of-mass 
velocity $v_c$, it only vanishes temporarily during rearrangement into the anti-Y-junction (green filled pentagon).

%DEFINITIONS SUMMARY\\
%fission (fw)    :$d > r_c		 						\ \& \ 							 \phi \leq 90^{\circ}$\\
%fission (bw)    :$d > r_c 								\ \& \  90^{\circ} < \phi < 180^{\circ}$\\
%two clusters    :$d > r_c 								\ \& \ 							 \phi=180^{\circ}$\\
%anti-Y-junction :$\alpha > 90^{\circ}    \ \& \ 180^{\circ} < \phi \leq 360^{\circ}$\\
%kite cluster    :$\alpha > 90^{\circ}    \ \& \  90^{\circ} < \phi < 180^{\circ}$\\
%Y-junction (bw) :$\alpha \leq 90^{\circ} \ \& \  90^{\circ} < \phi < 180^{\circ}$\\
%Y-junction (fw) :$\alpha \leq 90^{\circ} \ \& \ 							 \phi \leq 90^{\circ}$

\subsection{Compact cluster $(N=5)$}

\begin{figure*}[thbp]
\begin{center}
\includegraphics[clip=,width=2\columnwidth]{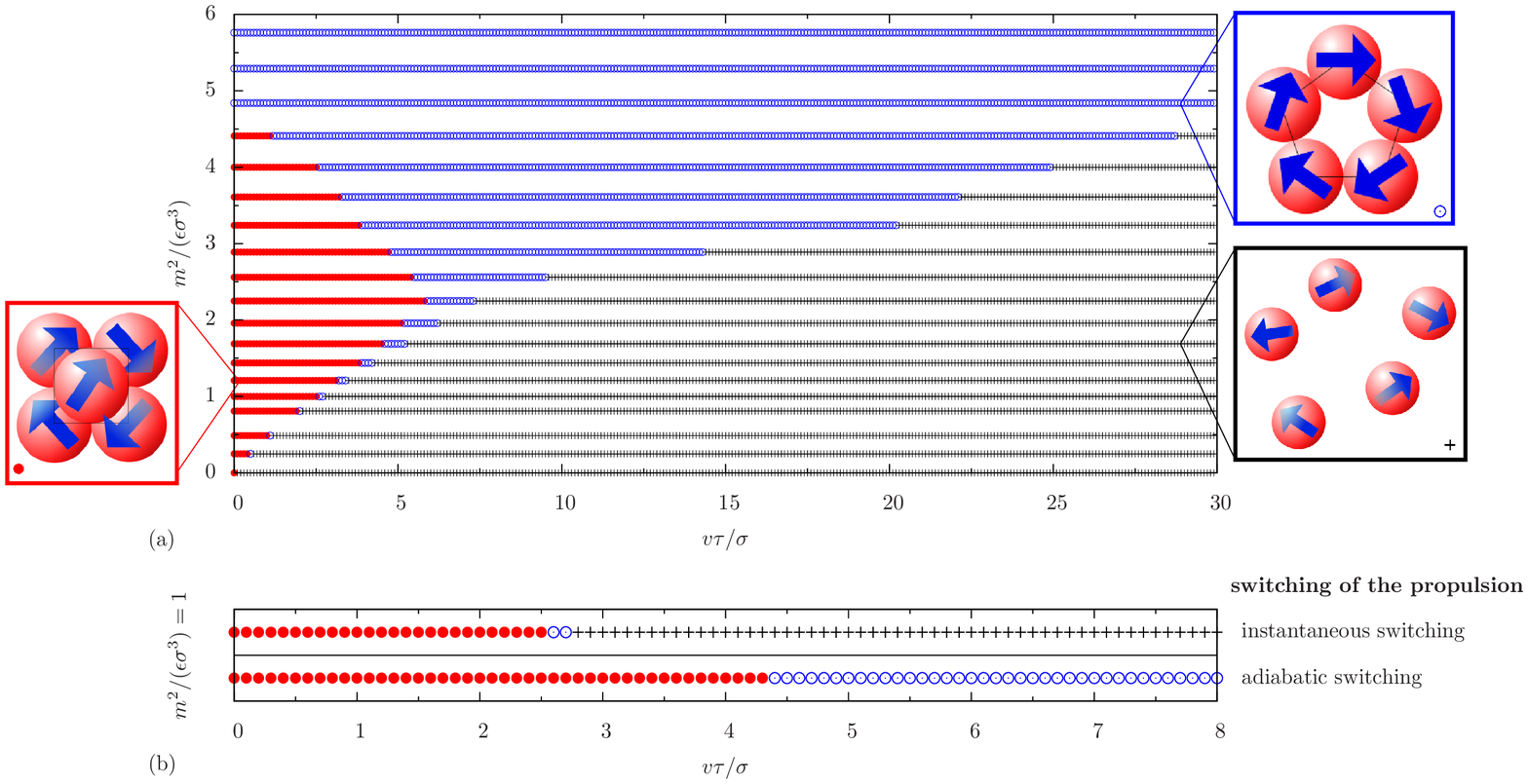}
\caption{ \label{PhaseDiag5er} (a) Emerging state diagram for an initially compact cluster of $N=5$ dipoles for 
varied reduced dipole strength $m^2/(\epsilon\sigma^3)$ and reduced self-propulsion velocity $v\tau/\sigma$. 
As in Fig.~\ref{PhaseDiag}, each symbol represents a type of final cluster configuration, 
shown through snapshots. (b) Comparison between instantaneous switching and adiabatic switching of the self-propulsion velocity 
$v$ for fixed dipole strength $m^2/(\epsilon\sigma^3)=1$.}
\end{center}
\end{figure*}

\begin{figure}[thbp]
\begin{center}
\includegraphics[clip=,width=1\columnwidth]{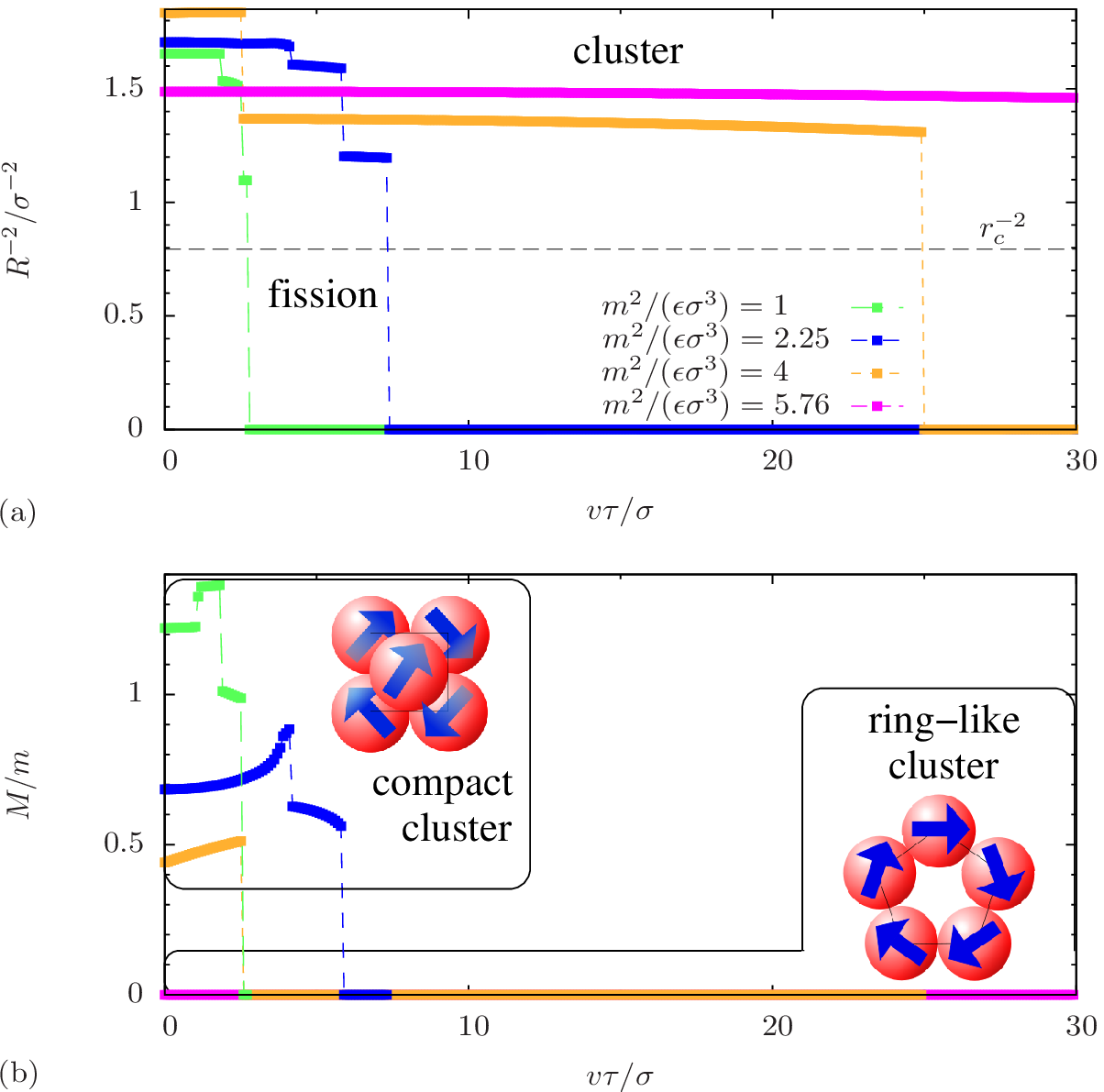}
\caption{ \label{DipMom} (a) Inverse squared radius of gyration $1/R^2$ and (b) reduced total dipole moment $M/m$ 
of an active cluster of $N=5$ dipoles for fixed dipole strength $m^2$ and varied self-propulsion velocity $v$. The dashed line
in (a) shows the inverse squared cut-off radius of the WCA potential $1/r_c^2$.}
\end{center}
\end{figure}

We finally study an initial three-dimensional cluster with a nonvanishing dipole moment 
for $N=5$ dipoles. Such a compact cluster of passive dipoles is metastable only
for $m^2/(\epsilon\sigma^3) < 4.5$, see Fig.~\ref{PhaseDiag5er}(a). For $m^2/(\epsilon\sigma^3) > 4.5$ this cluster will spontaneously 
transform into a ring-like structure which is the corresponding ground state~\cite{Miller}.
If self-propulsion is turned on instantaneously, this cluster will either stay stable and move
 on a helical trajectory as discussed in Sec.~\ref{Sec:Trajectories} (for $m^2/(\epsilon\sigma^3) < 4.5$) 
or spontaneously transform into a ring (for $m^2/(\epsilon\sigma^3) > 4.5$). 

The emerging state diagram for the final cluster is presented in Fig.~\ref{PhaseDiag5er}(a).
On top of the helical cluster motion and the ring-like structure, fission occurs at high self-propulsion $v$.
As a function of increasing $v$, the compact cluster first transforms into a ring before it splits apart.
There is a critical drive $v=v_c \approx 6 \sigma/\tau $ beyond which the compact initial cluster looses its stability.
For fixed drive $v<v_c$, an interesting reentrant scenario
for increasing dipolar strength $m$ occurs.
There is fission at small $m$, then the ring-structure is emerging as a final state, subsequently the compact cluster
is the final state and then the ring structure is getting stable again.

The comparison of different protocols is shown in
Fig.~\ref{PhaseDiag5er}(b). In qualitative accordance with the case of $N=4$ particles 
discussed above there are fewer (in this case only two) final states.
The fission is missing completely for adiabatic switching and the final cluster conformation for large 
self-propulsion velocities $v$ is a two-dimensional ring.

We have documented the fission of the cluster by monitoring the cluster radius $R$ of gyration defined via
\begin{eqnarray}
R^2= \frac{1}{N} \sum_{i=1}^{N} \left(\br_i - {\bf R}_c\right)^2\,.
\end{eqnarray}
Results for $R$ for an instantaneous switching protocol are shown in Fig.~\ref{DipMom}(a). 
Here the inverse of $R^2$ is plotted such that a cluster explosion is indicated by a vanishing $1/R^2$.
The dashed line indicates a size comparable to the cut-off distance $r_c$.
A comparison of $R$ relative to this line clearly reveals the 
transition from a compact cluster towards fission.
For completeness we have also shown the cluster magnetization $M$ in Fig.~\ref{DipMom}(b)
which is a good order parameter for a ring-like structure as $M$ vanishes there. These results
confirm the full phase diagram of Fig.~\ref{PhaseDiag5er}(a).

\section{Conclusion}
\label{Sec:conc}

We have studied the dynamical response of initially passive and metastable soft-sphere dipole clusters to the 
onset of an internal self-propulsion.
The latter is modelled by an internal effective force along the dipole moment
and the dynamics is completely overdamped in a solvent at low Reynolds number. Even though the cluster is small,
a wealth of different types of motion is obtained, which depend on the interaction parameters, the strength of the 
self-propulsion, and the self-propulsion switch-on protocol.
Moving cluster structures emerge which are not stable
in equilibrium. Interestingly, if the self-propulsion is very large and applied quickly (``instantaneous switching''), the cluster
shows a permanent fission but there is no fission if the drive is applied slowly (``adiabatic switching'').
 The center-of-mass of the cluster moves on linear, circular, or
 helical paths, depending on the initial magnetization of the cluster.

Self-propelled dipolar particles can be realized as Janus particles~\cite{Kretzschmar}, 
with an embedded dipole moment~\cite{SaguesPRL08,JanusDielectric,MagneticJanusColloids,2012GranickNature} where the 
self-propulsion \cite{Baraban_Nano2013,Baraban_ACSNano2013} is generated by self-diffusiophoresis. 
Cluster aggregation has indeed been observed~\cite{LB_pc}.
Furthermore, dipolar particles driven by external magnetic fields are conceivable as experimental
 realizations at microscopic~\cite{snezhko_prl} as well as macroscopic length scales~\cite{Grzybowski2000}. These particles 
can readily be observed in real-space such that our predictions are, in principle, verifiable.
Another realization of self-propelled dipoles are active droplets which are filled with a liquid crystal. 
These possess an inner topological defect, shifted from their center
of mass, which induces an electric dipole moment. Here, the strength of the dipole moment is proportional to the self-propulsion, which may
lead to new interesting effects~\cite{1988Lavrentovich,2014Herminghaus}.

Our work can and should be extended towards several directions for future research. First of all,
soft spheres with a non-central dipole moment have been considered recently in the passive case and 
the stable cluster structure was found to be different to that of central dipoles~\cite{Kantorovic,Holm2}. 
Moreover, one can imagine that the two directions of dipole moment and self-propulsion are not collinear 
which is expected to lead to even more complex dynamics. In this case, fission is expected also for slow switching.
For microswimmers, Brownian fluctuations induced by the solvent needs to be considered and incorporated into the 
dynamics~\cite{SvT2008,WittkowskiPRE12}.

A more detailed modelling which we shall pursue in the future is an inclusion of hydrodynamic
interactions by using more complicated friction tensors. The friction tensor
corresponding to the direct forces (or body forces) is derived from the pairwise interparticle potential. It
can be treated by the Oseen or Rotne-Prager tensor \cite{PuseyLH,RexLowenPRL08,Dhontbook}, as would be done
 for passive particles. On this level, strictly speaking a dumbbell has another 
friction tensor than of two spheres.
It is important to note that these friction tensors do not affect 
the initial equilibrium structure of the cluster. The swimming process, however, needs to be modelled
with a tensor that decays faster with interparticle distance than the Oseen tensor 
~\cite{Ramaswamy02,Ramaswamy04}. We expect that solving the coupled equation of motion
could give rise to new unexpected cluster dynamics as both interactions 
(the hydrodynamic and the dipolar one) are long-ranged and therefore compete. Moreover, the behaviour
will depend on the hydrodynamic boundary conditions. An unbounded solvent around the cluster will be described
by a different friction tensor than the motion of dipolar particles on a substrate~\cite{LaugaPRL13} 
or on a pending  air-liquid interface which will make 
the full problem even more complicated.
Finally, particle shapes different from that of a sphere can be studied
 like C- or L-shaped particles~\cite{BtHNatComm,Kaiser2carrier,WensinkPRE14}, 
which tend to perform a circular motion if they are self-propelled even in absence of any dipolar interaction.

\vspace{0.1cm}
\acknowledgments
We thank Eugene Terentjev and Peet Cremer for helpful discussions and Sandra Held for proofreading the manuscript.
This work was supported by the science priority program
SPP 1681 of the German Science Foundation (DFG) and by the ERC
Advanced Grant INTERCOCOS (Grant No. 267499).

\bibliography{refs}

\end{document}